\documentclass[10pt, final, conference]{IEEEtran}
\usepackage[utf8x]{inputenc}
\usepackage{graphicx}
\usepackage{footnote}
\usepackage[ dvips, 
  bookmarks = true,
  bookmarksopen = true,
  bookmarksnumbered = true,
  breaklinks = true,
  linktocpage,
  pagebackref = true,
  colorlinks = true,
  linkcolor = blue,
  urlcolor  = blue,
  citecolor = red,
  anchorcolor = green,
  hyperindex = true,
  hyperfigures
  ]{hyperref}

\title{Video Tester -- A multiple-metric framework for video quality assessment over IP networks}

\author{
  \authorblockN{Iñaki Ucar\authorrefmark{1}, Jorge Navarro-Ortiz\authorrefmark{2}, Pablo Ameigeiras\authorrefmark{2} and Juan M. Lopez-Soler\authorrefmark{2}}

  \authorblockA{\authorrefmark{1}
    Dept. of Automatics and Computer Science\\
    Public University of Navarra, 
    Campus Arrosadia, 31006 Pamplona, Spain\\
    i.ucar@unavarra.es
  }
  \authorblockA{\authorrefmark{2}
    Dept. of Signal Theory, Telematics and Communications\\
    University of Granada, 
    C/ Periodista Daniel Saucedo Aranda s/n, 18071 Granada, Spain\\
    \{jorgenavarro,pameigeiras,juanma\}@ugr.es
  }
}

\begin{document}

\maketitle

\begin{abstract}
  This paper presents an extensible and reusable framework which addresses the problem of video quality assessment over IP networks. The proposed tool (referred to as Video-Tester) supports raw uncompressed video encoding and decoding. It also includes different video over IP transmission methods (i.e.: RTP over UDP unicast and multicast, as well as RTP over TCP). In addition, it is furnished with a rich set of offline analysis capabilities. Video-Tester analysis includes QoS and bitstream parameters estimation (i.e.: bandwidth, packet inter-arrival time, jitter and loss rate, as well as GOP size and I-frame loss rate). Our design facilitates the integration of virtually any existing video quality metric thanks to the adopted Python-based modular approach. Video-Tester currently provides PSNR, SSIM, ITU-T G.1070 video quality metric, DIV and PSNR-based MOS estimations. In order to promote its use and extension, Video-Tester is open and publicly available.
\end{abstract}\vspace*{2mm}


\let\thefootnote\relax\footnote{
 \\\textcopyright 2012 IEEE. Personal use of this material is permitted. Permission from IEEE must be obtained for all other uses, in any current or future media, including reprinting/republishing this material for advertising or promotional purposes, creating new collective works, for resale or redistribution to servers or lists, or reuse of any copyrighted component of this work in other works.\\

 \noindent DOI: \hyperref{http://dx.doi.org/10.1109/BMSB.2012.6264243}{}{}{10.1109/BMSB.2012.6264243}
}

\section{Introduction}

During the last years, there has been a continuous growth of video traffic over packet-switched networks \cite{cisco}. Users' demand is determinant in the proliferation of a wide range of video-based services and applications such as video on demand (VoD), live streaming, video calling, video instant messaging, video monitoring, webcam traffic, Internet video to TV, live Internet TV, mobile video, etc. All these heterogeneous services are designed to fulfill different purposes and requirements \cite{simpson}. In spite of their dissimilarities, they all share a common need. Namely, to make use of tools for video quality evaluation.

The issue of video quality assessment (by defining the proper metrics) is of paramount importance for codec design, different transmission methods evaluation and network planning among others. Indeed, this problem has generated a lot of contributive research efforts \cite{chikkerur} along with the specification of several video quality measurement standards \cite{winkler2}.

However, public available tools that systematize the video quality assessment over IP networks are scarce. In this respect, it is also valuable that the designed assessment tool will be highly extensible and reusable since it eases the quick inclusion of any metric.

In this work, we propose Video-Tester, an open-source modular framework \cite{videotester} which provides a complete and flexible solution to the problem of video quality assessment over IP networks. Video quality evaluation is a multidimensional problem that requires to take into consideration multiple criteria, sometimes located at different levels. Our tool works at any of the three layers involved in any video over IP service \mbox{---the} bitstream level, the packet level and the upper picture level \cite{winkler1}---. As a remarkable benefit, Video-Tester is able to support any existing and future video quality metric.

The rest of the paper is organized as follows. Section II provides the state of the art of the addressed problem. Section III summarizes the proposed framework. It explains our Video-Tester design and also reviews the current supported metrics. Section IV outlines some experimental results that validate the proposed framework and, finally, Section V concludes the paper and discusses future work and extensions.

\section{Related work}

\begin{figure*}
 \centering
 \includegraphics[width=0.7\textwidth]{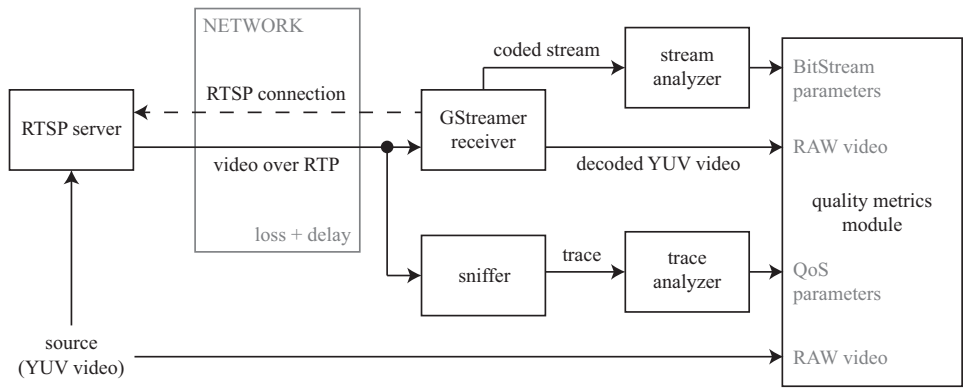}
 \caption{Operation scheme.}
 \label{scheme1}
\end{figure*}

Several quality evaluation toolboxes have been recently proposed, like the implemented by Murthy and Karam \cite{ivquest}. However, in general, these applications are mainly restricted to codec testing and metric comparisons, because they do not consider multimedia transmission methods and the degradation and quality impact derived from this process.

On the other hand, it is also remarkable the seminal work (EvalVid) proposed by Klaue \textit{et al.} \cite{evalvid2}, a widely used tool-set for video transmission evaluation. This tool-set does actually include transmission, some quality of service (QoS) parameter estimation (packet loss, delay and jitter) and some bitstream parameter evaluations (frame loss rate and frame jitter). EvalVid video quality evaluation techniques include PSNR, Structural Similarity index (SSIM) \cite{ssim}, Mean Opinion Score (MOS, mapped from PSNR) and Distortion in Inverval (DIV, mapped from MOS as described in \cite{evalvid3}).

However, this is a multi-tool framework that requires 3rd party tools like \textit{tcpdump} (packet capture), \textit{ffmpeg} (video encoding and decoding), \textit{netcat} and \textit{MP4Box} \cite{gpac} (video multiplexation). This fact makes the automatization of its use a bit difficult. Furthermore, the use of MP4 multimedia containers restricts the available codecs. Added to that, the election of ISO-C programming language is definitively appropriate  for code efficiency (not so for porting because of the necessary 3rd party tools), but if new quality metrics are needed, the source code must be modified.

Additionally, a common impairment in video transmission evaluation is related to frame losses. The decoded RAW video at the receiver end will have fewer frames than the original one if packet losses occur. As a consequence, we have to deal with frame misalignment in order to apply picture metrics like PSNR. In \cite{evalvid2}, it is stated that Fix-Video (FV) tool solves this problem. Nevertheless, to the author's knowledge the actual version of this framework \cite{evalvid1} does not include the FV tool.

\section{Proposed framework}

Video-Tester comprises a single command-line application that works with a number of configuration files that ease its script-based execution. Once Video-Tester is launched at the two particular end-points, one side acts as the server and the other one as the client. The client side can optionally be launched with a user-friendly graphical user interface (GUI) in order to plot metrics automatically.

Video-Tester is written in Python in order to promote its extensibility. Video processing (specifically encoding, decoding and transmission) is performed by using the valuable GStreamer library \cite{gstreamer} due to its broad and sustained support by the research community. As a beneficial consequence, the codec support is subjected to the GStreamer support and, hence, it is not constrained to a given container format.

\subsection{Video-Tester design and operation}

Figure \ref{scheme1} shows the general operation scheme. The server side is concurrent and talks to clients using the XML-RPC protocol. Both the server and the client are able to work behind a NAT router (with port redirection at the server side). Each client is able to select the following parameters in order to start a new test:

\begin{itemize}
 \item The uncompressed video file to transmit, selected from a common local database (Video-Tester accepts lossless encoded videos like \cite{evalvid4}).
 \item The codec (H.263 \cite{h263}, H.264 \cite{h264}, MPEG-4 Visual \cite{mpeg42} and Theora \cite{theora} are currently available).
 \item The bitrate (not all bitrates are available with some codecs).
 \item The frame rate.
 \item The transmission method (RTP over UDP unicast or multicast, and RTP over TCP are supported).
\end{itemize}

Then, the server side sets up a GStreamer-based RTSP server (see \cite{gstrtspserver}) instance with the above parameters. At the client side, a GStreamer-based client receives the video signal. Remarkably, it is programmed to keep the frame rate constant. That is, if transmission losses cause the frame rate to fall, the receiver end duplicates properly the available frames.

At the same time and executed in a separate thread, the packet-sniffer saves concurrently a network trace. So that whenever the transmission is completed, the following files are provided:

\begin{itemize}
 \item The network trace (PCAP file \cite{pcap}).
 \item Received compressed video (probably impaired with losses).
 \item Received uncompressed video (probably impaired with losses).
 \item Reference video (from the local database).
\end{itemize}

At this point, the client side performs the offline analysis. From the RTP trace, we extract the following QoS parameters:

\begin{itemize}
 \item Packet size.
 \item RTP sequence number.
 \item RTP timestamp.
 \item Packet arrival time.
\end{itemize}

From the coded stream, we extract the following bitstream parameters:

\begin{itemize}
 \item Frame type (I, P, B).
 \item Frame size.
\end{itemize}

\begin{figure}
 \centering
 \includegraphics[width=0.6\linewidth]{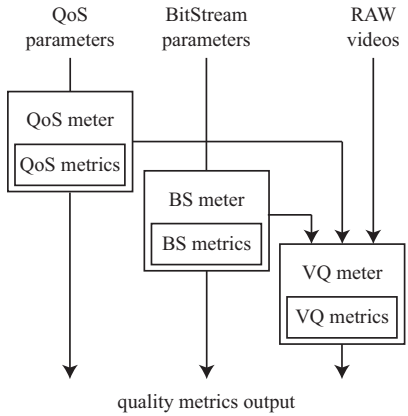}
 \caption{Quality metrics module.}
 \label{scheme2}
\end{figure}

The quality metrics module receives all this information. Figure \ref{scheme2} shows a detailed overview of this module. It contains three submodules called \textit{meters} that isolate the implemented metrics from the rest of the application logic. Meters keep track of the available metrics and they communicate with those metrics through a standard interface. Therefore, implementing a new metric is as easy as writing a new Python class and registering it at the proper meter, without modifying the application core.

\subsection{Supported metrics}

The \emph{QoS submodule} provides the following metrics:

\begin{itemize}
  \item Latency ($L$), averaged from several ($N$) round-trip time ($RTT$) values.
  \begin{equation}
    L = \frac{1}{N}\sum_{n=1}^{N}{\frac{RTT(n)}{2}}
  \end{equation}

  \item Packet inter-arrival time ($\Delta$), where $R_i$ is the arrival time for $i$-th packet.
  \begin{equation}
    \Delta(i) = R_i - R_{i-1}
  \end{equation}

  \item Jitter ($J$), as described in \cite{rtp}, where $R_i, R_j$ are arrival times and $S_i, S_j$ are RTP timestamps for packets $i, j$.
  \setlength{\arraycolsep}{0.14 em}
  \begin{eqnarray}
    D(i,j) &=& (R_j - R_i) - (S_j - S_i) \nonumber \\
    &=& (R_j - S_j) - (R_i - S_i) \\
    J(i) &=& J(i-1) + \frac{|D(i-1,i)| - J(i-1)}{16}
  \end{eqnarray}
  \setlength{\arraycolsep}{5pt}

  \item Clock skew ($T$), the relative offset between packet arrival time and RTP timestamp, where $R_i$ is the arrival time and $S_i$ is the RTP timestamp for packet $i$.
  \begin{equation}
    T(i) = S_i - R_i
  \end{equation}

  \item Instantaneous bandwidth ($B$) for packet $i$, where $Size_n$ is the size of packet $n$ and $N$ is the number of packets in the last second.
  \begin{equation}
    B(i) = \sum_{n=i-N}^i{Size_n}
  \end{equation}

  \item Packet loss rate ($PLR$), where $Seq_n$ is the RTP sequence number of packet $n$ and $N$ is the total number of packets.
  \begin{equation}\label{plr}
    PLR = \frac{1}{N} \sum_{n=2}^N{Seq_n-(Seq_{n-1}+1)}
  \end{equation}

  \item Packet loss distribution ($PLD$). If we divide the transmission time in $K$ intervals, we can apply (\ref{plr}) to every interval $k$.
  \begin{equation}
    PLD(k) = \frac{1}{N_k} \sum_{n=2}^{N_k}{Seq_{n,k}-(Seq_{n-1,k}+1)}
  \end{equation}
\end{itemize}

The \emph{bitstream (BS) submodule} provides the following metrics:

\begin{itemize}
  \item Video stream framing structure (at the receiver end, probably impaired with frame losses).
  \item Reference video stream framing structure (with no losses).
  \item Group of pictures (GOP) size, averaged from measured GOP sizes once atypical values has been discarded.
  \setlength{\arraycolsep}{0.14 em}
  \begin{eqnarray}
    &&GOP size = E[X] \\
    \hbox{with} \quad X = \{&GOP_n& : \nonumber \\ &GOP_n& \in [E[GOP]-\sigma, E[GOP] + \sigma ] \} \nonumber
  \end{eqnarray}
  \setlength{\arraycolsep}{5pt}

  \item I-frame loss rate. We count an I-frame loss when
  \begin{equation}
    GOP_n > E[GOP] + \sigma
  \end{equation}
\end{itemize}

The \emph{video quality (VQ) submodule} provides the following metrics:

\begin{itemize}
  \item PSNR.
  \item SSIM index \cite{ssim}.
  \item MOS, mapped from PSNR \cite{evalvid3}.
  \item DIV, mapped from MOS (as proposed in \cite{evalvid3}).
  \item As an illustration, we provide ITU-T G.1070 video quality metric \cite{g1070}, because it provides an example of gathering information from other submodules, specifically the packet loss rate.
\end{itemize}

\section{Experimental results}

\subsection{Frame misalignment}

\begin{figure}
 \centering
 \includegraphics[width=\linewidth]{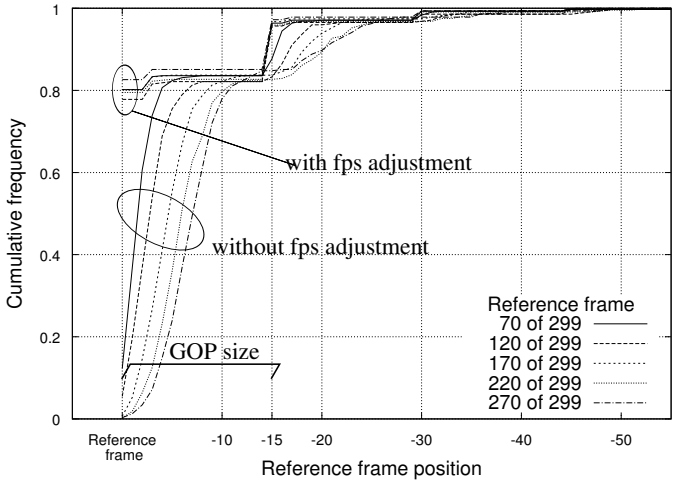}
 \caption{Frame misalignment test.}
 \label{misalignment1}
\end{figure}

\begin{figure}
 \centering
 \includegraphics[width=\linewidth]{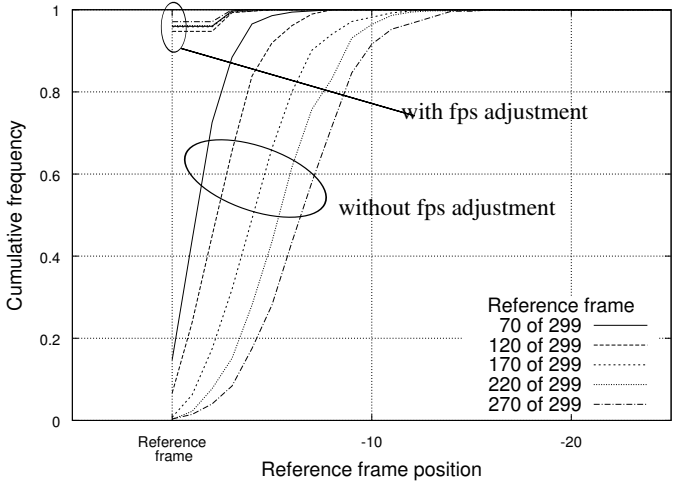}
 \caption{Frame misalignment test (once the videos without the first frame were removed).}
 \label{misalignment2}
\end{figure}

To test the effectiveness of the Video-Tester frame rate sustainer procedure, a particular scenario with 2 \% of packet loss probability was chosen. The video \texttt{akiyo\_cif.264} \cite{evalvid4} was encoded with H.263 (GOP size of 15) at 128 kpbs and 25 fps, and transmitted over UDP unicast. 600 video transmissions were performed. For each transmission, we get the received video with and without the Video-Tester frame rate sustainer element.

We choose 5 frame positions (70, 120, 170, 220 and 270 of 299 total frames) from the reference video in order to compare misalignments through all the tests. Figure \ref{misalignment1} shows the cumulative frequency of the position occupied by the reference frame. The received video presents severe misalignments in absence of the frame rate sustainer. Instead, our method highly increases the probability that the frames are aligned. However, note that there is an anomaly at position $-15$.

This anomaly can be explained attending to the GOP size. When the loss occurs at the first frame of the first GOP (an I-frame), the remaining frames (P or B) have no reference to be decoded. Therefore, the loss of this first frame implies the loss of the whole GOP. To the receiver, the video transmission starts one GOP later, so the frame rate sustainer has no work. This point can be proved with Figure \ref{misalignment2}. It shows the same analysis, once the videos without the first frame were removed.

Accordingly, in order to do reliable Full-Reference calculations (like PSNR) with our frame adjustment method, we can dismiss the first GOP if the received video lacks the first frame.

\subsection{Exemplary tests}

A bank of tests was designed in order to show the kind of results that can be achieved with our Video-Tester tool. The video \texttt{akiyo\_cif.264} \cite{evalvid4} was encoded with H.263 at 300 kpbs and 25 fps, and transmitted over UDP unicast. 10 tests were launched with increasing packet loss probability, from 0.1 \% to 3.7 \%. Within each test, we measured the I-frame loss rate and the percentage of frames with certain level of MOS.

\begin{figure}
 \centering
 \includegraphics[width=\linewidth]{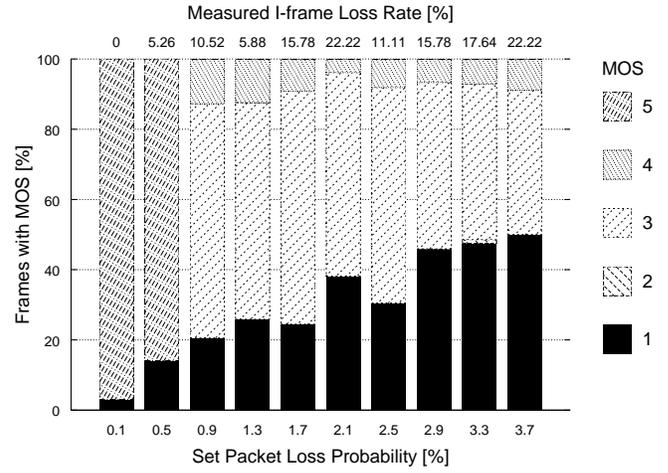}
 \vspace*{-2mm}\caption{MOS ratings measured in scenarios with different packet loss probability.}
 \label{mos}
\end{figure}

Results are shown in Figure \ref{mos}. Each stacked bar belongs to one test. The bottom x-axis shows that tests are ordered from left to right with increasing packet loss probability. The top x-axis shows the measured I-frame loss rate for each test. As expected, we can observe a progressive deterioration of both I-frame loss rate and MOS ratings.

\section{Conclusion}

This paper proposes Video-Tester, an extensible and reusable single framework for video quality assessment over IP networks. On the basis of two end-points (public or private IP addresses), it comprises all the procedures involved in video over IP communications. More specifically, it accounts from capturing to rendering video, which involves encoding, sending, receiving and decoding the video signal.

Video-Tester estimates the proper parameters and characteristics required for the overall system evaluation, and due to its modularity, it can evaluate the impact of any of the components ---at any level--- on the final video quality.

Notably, Video-Tester is a Python-based tool and, hence, it benefits from Python key distinguishing features. It also uses the valuable GStreamer library \cite{gstreamer}. More precisely, to achieve the greatest flexibility and for making it maximally configurable, a GStreamer RTSP server \cite{gstrtspserver} is designed. Moreover, a procedure to solve the frame misalignment problem in lossy scenarios is proposed.

Summing up, Video-Tester includes the main features of previous works and adds further improvements in terms of usability, extensibility, codec support, support of transmission methods and metric robustness (frame alignment) in case of losses. The wide range of extracted parameters allows the implementation of virtually any kind of video quality metric: Full-Reference, No-Reference or Reduced-Reference metrics; as well as any level picture-, packet-, bitstream-based or even hybrid metrics.

\section*{Acknowledgments}

This work was partially supported by the ``Ministerio de Ciencia e Innovación'' of Spain under research project TIN2010-20323.

\IEEEtriggeratref{8}
\bibliographystyle{IEEEtran}
\bibliography{references}

\end{document}